\documentclass{ptephy_v1}

\usepackage{boites}
\usepackage{amsmath,amssymb}
\usepackage{graphicx}
\usepackage{float}
\usepackage{tabularx}
\usepackage{xcolor}

\def\intt{\int \!\!...\!\! \int}
\def\<{\left\langle}
\def\>{\right\rangle}
\def\({\left(}
\def\){\right)}
\def\e{{\rm e}}

\begin{document}

\title{
Eigenphase distributions of unimodular circular ensembles
}

\author{\name{\fname{Shinsuke} \surname{Nishigaki}}{\ast}} 

\address{\affil{~}{Graduate School of Natural Science and Engineering, 
Shimane University, Matsue 690-8504, Japan}
\email{mochizuki@riko.shimane-u.ac.jp}
}

\begin{abstract}
Motivated by the study of Polyakov lines in gauge theories,
Hanada and Watanabe recently presented a conjectured formula for
the distribution of eigenphases of Haar-distributed random
 $\mathrm{SU}(N)$ matrices ($\beta=2$), supported by explicit examples at small $N$
and by numerical samplings at larger $N$.
In this letter, I spell out a concise proof of their formula,
and present its orthogonal and symplectic counterparts, i.e. the eigenphase distributions of
Haar-random unimodular
symmetric ($\beta=1$) and selfdual ($\beta=4$) unitary matrices parametrizing
$\mathrm{SU}(N)/\mathrm{SO}(N)$ and $\mathrm{SU}(2N)/\mathrm{Sp}(2N)$, respectively.
\end{abstract}

\subjectindex{B83, B86, A10, A13}

\maketitle

\section{Foreword and motivation}
This letter is inspired by a conjectured formula (117) in Ref.\,\cite{Hanada}, presented without a proof.
A main goal of that paper is to quantify the partial deconfinement of  lattice gauge theories
at finite temperature
in terms of the statistical distribution of the eigenphases of the Polyakov line
$P(\vec{n})=U_0(0,\vec{n})U_0(1,\vec{n})\cdots U_0(L_t-1,\vec{n})$, treated as random
SU($N$) matrices.
From the gauge-theory point of view, it is crucial to consider a simple group SU($N$)
rather than semi-simple U($N$), obviously because the running of the coupling constant
for each simple or Abelian factor of a gauge group is different.
This naturally led the authors of Ref.\,\cite{Hanada} to conjecture the 
eigenphase distribution of {\em Haar-distributed} random SU($N$) matrices,
i.e.  the circular unitary ensemble (CUE) with a unimodular constraint $\det U=1$.
Their formula is based upon explicit examples at $N=2, 3$ and numerical samplings
at larger $N$ \cite{Fasi}.

In the study of so-called fixed trace random matrix ensembles \cite{Rosenzweig},
typically, the sum of the squared eigenvalues
$\mathrm{tr}\, H^2=\sum_{j=1}^N \lambda_j^2$ of random $N\times N$ Hermitian matrices $H$ is 
constrained to a specific value.
Although this type, or more generic type ($\mathrm{tr}\, V(H)=\mathrm{const.}$\,\cite{Akemann}) 
of constraints
respects the $\mathrm{U}(N)$ invariance of the unconstrained ensemble,
additional interactions among multiple eigenvalues 
induced by the trace constraint destroy
the determinantal property of their correlation functions.
Due to this difficulty, 
one often had to be content with,
either a macroscopic large-$N$ limit by the Coulomb-gas method \cite{Akemann}
or asymptotic universality at $N\gg 1$ \cite{Gotze}, 
while subtleties in the local correlations of eigenvalues
still remain elusive.
 
In this letter, I dedicate my tiny contribution to the field of constrained
random matrices,
namely a proof of the aforementioned conjecture
on the density of the eigenphases of Haar-random
SU($N$) matrices.
My proof encompasses Dyson's Threefold Way \cite{Dyson3FW} all at once,
as it automatically provides the densities of the eigenphases of 
Haar-random symmetric SU($N$) matrices ($U=U^T$) parametrizing the quotient SU($N$)/SO($N$)
and selfdual SU($2N$) matrices ($U=U^D:=J U^T J^{-1}, J=i\sigma_2\otimes \mathbb{I}_N$)
parametrizing the quotient SU($2N$)/Sp($2N$),
i.e. the circular orthogonal and symplectic ensembles (COE, CSE)
with unimodular constraints.
It would be my pleasure if this letter will serve as useful appendix to Ref.\,\cite{Hanada}.

\section{Theorem and proof}
{\bf Theorem}\\
Let 
$\{\e^{i\theta_1}, \ldots, \e^{i\theta_{N-1}}, \e^{i\theta_{N}}(=\e^{-i(\theta_{1}+\cdots+\theta_{N-1})})\}$
be the set of $N$ eigenphases of either
SU($N$) matrices ($\beta=2$), 
symmetric SU($N$) matrices ($\beta=1$), or
selfdual SU($2N$) matrices ($\beta=4$)
that are Haar-distributed.
Then the probability density of these eigenphases
is given by\footnote{%
Excluding an exceptional case with $\beta=1$, $N=2$, 
for which $\rho_{1,2}(\theta)=|\sin\theta|/2$ trivially follows.}
\begin{align}
\rho_{\beta,N}(\theta)=\frac{N}{2\pi}\times
\left\{
\begin{array}{ll}
{\displaystyle 1-(-1)^N\frac{2}{N}\cos N\theta}
& (\beta=2)\\
{\displaystyle 1-(-1)^N
\frac{\sqrt{\pi}(N-1)!}{2^{N-1}\Gamma(N/2+3/2)\Gamma(N/2+1)}
\cos N\theta}
& (\beta=1)\\
{\displaystyle
1-(-1)^N\frac{(2N)!!}{(2N-1)!!N}\cos N\theta+\frac{2}{(2N-1)N} \cos 2N\theta}
& (\beta=4)
\end{array}
\right\} .
\label{theorem}
\end{align}

\noindent
{\bf Proof.}
The normalized joint distributions of $N$ eigenphases 
$\{\e^{i\theta_1}, \ldots, \e^{i\theta_{N}} \}$
of Haar-distributed U($N$) matrices ($\beta=2$),
symmetric U($N$) matrices ($\beta=1$), and
selfdual U($2N$) matrices ($\beta=4$)
[denoted as C$\beta$E$(N)$, respectively] are well-known to be \cite{Dyson}
\begin{align}
d\mu_{\mathrm{C}\beta\mathrm{E}(N)}(\theta_1,\ldots,\theta_N)=
\frac{1}{C_{\beta,N}} \prod_{j=1}^N \frac{d\theta_j}{2\pi}\cdot |\Delta_N(\vec{\theta})|^\beta,
\ \ \ C_{\beta, N}=\frac{\Gamma(\beta N/2+1)}{\Gamma(\beta/2+1)^N}.
\end{align}
Here $\Delta_N(\vec{\theta}):=\prod_{1\leq j<k\leq N}(\e^{i\theta_j}-\e^{i\theta_k})$
stands for the Vandermonde determinant.
Upon imposing the unimodular constraint $\det U=\prod_{j=1}^N \e^{i\theta_j}=1$,
the joint distribution of $(N-1)$ {\em independent} eigenphases is given by
\begin{align}
d\mu_{\beta,N}(\theta_1,\ldots,\theta_{N-1})&=
\frac{1}{C_{\beta,N}}\prod_{j=1}^{N-1} 
\left.
\frac{d\theta_j}{2\pi}\cdot 
|\Delta_N(\vec{\theta})|^\beta
\right|_{\theta_N=-\sum_{j=1}^{N-1}\theta_j}
\nonumber\\
&=
\frac{1}{C_{\beta,N}}\int_{\theta_N} \prod_{j=1}^{N} 
\frac{d\theta_j}{2\pi}\cdot 
|\Delta_N(\vec{\theta})|^\beta
\cdot
2\pi\delta\Bigl(\sum_{k=1}^N \theta_k\ (\mbox{mod}\ 2\pi)\Bigr)
\nonumber\\
&=
\sum_{n=-\infty}^\infty  
\int_{\theta_N}
d\mu_{\mathrm{C}\beta\mathrm{E}(N)}(\theta_1,\ldots,\theta_N)
\prod_{k=1}^N\e^{in \theta_k}.
\label{dmuSUN}
\end{align}
Here $\int_{\theta_N}$ denotes an integral $\int_{-\pi}^{\pi}$ over the variable $\theta_N$,
and use is made of the Fourier expansion of the periodic delta function, 
$\delta(\theta \ \mbox{mod}\ 2\pi )=(2\pi)^{-1}\sum_{n} \e^{in\theta}$.

The probability distribution of a single eigenphase of unimodular matrices $U$ is
\begin{align}
\rho_{\beta,N}(\theta)&=
\mathbb{E}[\mathrm{tr}\, \delta(\theta+i \log U)]
\nonumber\\
&=
\intt_{-\pi}^\pi
d\mu_{\beta,N}(\theta_1,\ldots,\theta_{N-1})
\left.
\sum_{j=1}^N
 \delta(\theta-\theta_j)\right|_{\theta_N=-\sum_{j=1}^{N-1}\theta_j}
\nonumber\\
&=
 \sum_{n=-\infty}^\infty  
\intt_{-\pi}^\pi
d\mu_{\mathrm{C}\beta\mathrm{E}(N)}(\theta_1,\ldots,\theta_N)
\prod_{k=1}^N\e^{in \theta_k}
\cdot
N\delta(\theta-\theta_N).
\end{align}
We used the permutation symmetry of $\theta_j$'s.
After performing an integration over $\theta_N$ and 
a constant shift of the variables $\theta_j\mapsto\theta_j+\theta$
($j=1,\ldots, N-1$), 
it reads
\begin{align}
\rho_{\beta,N}(\theta)
&=
\frac{N}{2\pi}\frac{1}{C_{\beta,N}}
 \sum_{n=-\infty}^\infty  
\e^{inN\theta}
\intt_{-\pi}^\pi
\prod_{j=1}^{N-1} \(\frac{d\theta_j}{2\pi} \e^{in \theta_j} 
|1-\e^{i\theta_j}|^\beta
\)
|\Delta_{N-1}(\vec{\theta})|^\beta.
\label{sumn}
\end{align}
The integral in (\ref{sumn}) is renowned as
the Selberg integral \cite{Selberg}
in Morris's trigonometric form \cite{Morris} (see Ref.\,\cite{Forrester}, p.134),
\begin{align}
&\intt_{-\pi}^\pi
\prod_{j=1}^{N} \(\frac{d\theta_j}{2\pi} \e^{i\frac{a-b}{2} \theta_j} 
|1-\e^{i\theta_j}|^{a+b}
\)
|\Delta_{N}(\vec{\theta})|^{2\lambda}
\nonumber\\
&= 
(-1)^{\frac{a-b}{2}N}
\prod_{j=0}^{N-1}
\frac{
\Gamma(\lambda j+a+b+1)\Gamma(\lambda j+\lambda+1)
}{
\Gamma(\lambda j+a+1)\Gamma(\lambda j+b+1)\Gamma(\lambda+1)
}.
\label{Selberg}
\end{align}
Upon substituting $N\mapsto N-1,\; a=\beta/2+n,\; b=\beta/2-n,\; \lambda=\beta/2$
into (\ref{Selberg}), 
the LHS matches the integral in (\ref{sumn}) and 
the RHS is equal to
\begin{align}
\left\{
\begin{array}{ll}
{\displaystyle
N!\, \delta_{n,0}+(-1)^{N-1} (N-1)! \,\delta_{n,\pm1}
} & (\beta=2)\\
{\displaystyle
\frac{\Gamma(N/2+1)}{\Gamma(3/2)^N}\delta_{n,0}+
(-1)^{N-1}
\frac{(N-1)!}{\Gamma(N/2+3/2)\Gamma(1/2)^{N-1}}
\delta_{n,\pm1} }
& (\beta=1)\\
{\displaystyle
\frac{(2N)!}{2^{N}} \delta_{n,0}+(-2)^{N-1} N!(N-1)! \delta_{n,\pm1}+\frac{(2N-2)!}{2^{N-1}} \delta_{n,\pm2}
}
& (\beta=4)
\end{array}
\right. ,
\label{res}
\end{align}
except for a special case $\beta=1, N=2$.
Substitution of (\ref{res}) into (\ref{sumn}) yields
Theorem (\ref{theorem}). $\square$\\

Theorem (\ref{theorem}) for $\beta=2$ was conjectured in Ref.\,\cite{Hanada}, Eq.\,(117).
To the best of our knowledge,
either a proof or even a conjecture of Theorem for $\beta=1$ and 4
has not been spotted anywhere in the literature.
The above procedure is obviously applicable for imposing a constraint $\sum_{j=1}^N \theta_j=0$
on the circular $\beta$-ensemble involving $|\Delta_{N}(\vec{\theta})|^\beta$
at a generic integer $\beta$.

\section*{Acknowledgment}
This work is supported 
by a JSPS Grant-in-Aid for Scientific Research (C) No.\,7K05416.

\end{document}